\documentclass[nofootinbib, reprint, amsmath,amssymb, aps,]{revtex4-1}
\usepackage{graphicx}
\usepackage{dcolumn}
\usepackage{bm}
\usepackage[colorlinks]{hyperref}

\usepackage{graphicx,times}
\usepackage{subfigure}
\usepackage{amsmath}
\usepackage{mathrsfs}
\usepackage{cases}
\usepackage{bm}
\usepackage{longtable}
\usepackage{hyperref}
\usepackage{epstopdf}
\usepackage{amsmath,bm}
\usepackage{amssymb}
\usepackage{natbib}
\usepackage{morefloats}
\usepackage{multirow}
\usepackage{array}
\usepackage{verbatim}
\usepackage[stable]{footmisc}
\usepackage{stackrel}
\usepackage{graphicx}
\usepackage{dcolumn}
\usepackage{bm}

\newcommand{\bX}{{\bf X}}

\newcommand{\br}{{\bf r}}
\newcommand{\by}{{\bf y}}
\newcommand{\bx}{{\bf x}}

\newcommand{\bxi}{{\mbox{\boldmath $\xi$}}}

\begin{document}


\title{Topological theory of physical fields}

\author{Amir Jafari}
 \email{elenceq@jhu.edu}%
\affiliation{%
Johns Hopkins University, Baltimore, USA\\
}%

\author{Ethan Vishniac}
 \email{evishni1@jhu.edu}
\affiliation{%
Johns Hopkins University, Baltimore, USA\\
}%

\date{December 18, 2020}

\begin{abstract}

We study the topology associated with physical vector and scalar fields. A mathematical object, e.g., a ball, can be continuously deformed, without tearing or gluing, to make other topologically equivalent objects, e.g., a cube or a solid disk. If tearing or gluing get involved, i.e., the deformation is not continuous anymore, the initial topology will consequently change giving rise to a topologically distinct object, e.g., a torus. This simple concept in general topology may be employed in the study of physical systems described by fields. Instead of continuously deforming objects, we can take a continuously evolving field, with an appropriately defined topology, such that the topology remains unchanged in time unless the system undergoes an important physical change, e.g., a transition to a different energy state. For instance, a sudden change in the magnetic topology in an energetically relaxing plasma, a process called reconnection, strongly affects the dynamics, e.g., it is involved in launching solar flares and generating large scale magnetic fields in astrophysical objects. In this topological formalism, the magnetic topology in a plasma can spontaneously change due to the presence of dissipative terms in the induction equation which break its time symmetry. We define a topology for the vector field $\bf F$ in the phase space $(\bf x, F)$. As for scalar fields represented by a perfect fluid, e.g., the inhomogeneous inflaton or Higgs fields, the fluid velocity $\bf u$ defines the corresponding topology. The vector field topology in its corresponding phase space $(\bf x, F)$ will be preserved in time if certain conditions including time reversal invariance are satisfied by the field and its governing differential equation. Otherwise, the field's topology can suddenly change at some point, similar to a spontaneously broken symmetry, as time advances, e.g., corresponding to an energy transition. 
\end{abstract}

\pacs{Valid PACS appear here}

\maketitle


\section{Introduction}

Vector and scalar fields are functions of space and time which  assign, respectively, a vector or a scalar to each point in space. It goes without saying that these fields, and the obvious generalization to tensor fields, play a very important role in different branches of physics. Familiar examples of vector fields include magnetic and electric fields while gravitational potential and the Higgs field exemplify scalar fields. Tensor fields, such as Riemann curvature tensor, are likewise commonplace tools, e.g., in general relativity. Such physical fields are usually governed by differential equations, like Maxwell's equations, whose solutions provide us with local information about the corresponding system. However, it is interesting, and perhaps useful, to think about the overall shape or global topology of a given field. The magnetic field in a plasma gives us valuable information about the local dynamics, for instance in terms of the evolution of sunspots or coronal mass ejections on the sun. But we can go further and ask if the topology of the solar magnetic fields in a global sense, once appropriately defined, gives us any additional and useful information, e.g., regarding eruptive changes in the field direction and energy which drive fast jets of hot plasma. 

It may seem that an intuitive notion of topology is associated with vector fields, which are usually visualized in terms of their field lines (aka integral curves or streamlines). Yet, vector field topology and topology change are seldom given a precise operational meaning in physical applications. On the other hand, as it turns out, the notion of field lines and also that of field's topology are not trivial at all, in particular, in the context of time-dependent, stochastic fields in dissipative media. Magnetic and velocity fields are perhaps among the most frequently encountered vector fields in physics, hence they will be employed in this paper as examples to represent different concepts. For instance, the magnetic field threading an electrically conducting fluid such as a plasma is said to have a certain topology that may change spontaneously; a process referred to as magnetic reconnection which may strongly affect the system's dynamics, e.g., it may be the underlying process launching solar flares. In fact, electric field topology (\cite{Heikkila1978}; \cite{Goertzetal1979}), magnetic field topology (\cite{Stenzeletal1981}; \cite{Titovetal1999}) and velocity field topology (see e.g., \cite{Moffatt1985}; \cite{Helman1991}) are commonly used terminologies in different fields of research. In many problems, for example the generation and reconnection of astrophysical magnetic fields, the field topology plays a crucial role (see e.g., \cite{Lietal2007}; \cite{Schindleretal1988}; \cite{Eyink2015}; \cite{JafariandVishniac2018}). However, it is not clear, from a mathematical point of view, what is exactly meant by magnetic topology change in such dynamic, diffusive and chaotic environments.

The mathematical field of topology studies those properties of geometric objects that are preserved under continuous deformations, i.e., stretching, twisting and bending but not tearing or gluing. Topology is a structure that helps us define, rigorously, the notion of deformation and continuity. A topological space is any arbitrary set that is equipped with a topology. For instance, any Euclidean space with any dimension is a set of points between any pair of which a notion of distance is defined. This notion of distance, formally called a metric, helps define continuity and deformation, therefore, it allows to define a metric topology for the space. Suppose a Euclidean space $X$ is continuously deformed to space $Y$: the deformation is a map that takes every point in $X$ to only one point in $Y$ and vice versa, because $Y$ can be deformed back to $X$. Thus, the map should be onto and one-to-one. It is a continuous map since we wish to avoid tearing/gluing, i.e., arbitrarily close points in $X$ must remain arbitrarily close in $Y$. This is the very definition of a continuous map $f:X\rightarrow Y$. Symmetry requires the inverse map $f^{-1}$ to be continuous as well. Therefore, the continuous deformations used in topology are mathematically defined as one-to-one, onto and continuous maps with continuous inverse between topological spaces---homeomorphisms. Any property invariant under homeomorphisms is a topological property. 

Hence, homeomorphisms are continuous maps between topological spaces that keep the topological properties intact; if a time dependent vector field has a preserved topology, its time translation must be a homeomorphism, i.e., it should continuously take the field at time $t_0$ and map it to another field with the same topology at a later time $t_1>t_0$. The topology of an object such as a ball is well-defined and easy to visualize. How do we define a topology associated with a vector field? How do we ensure that the field's time translation is a homeomorphism, thus its topology is preserved in time? Under what conditions, can the vector field topology change to let, for example, magnetic reconnection occur in a magnetized plasma, producing eruptive jets of fluid like what we observe on the solar surface? What about scalar fields; can we take a scalar field such as the Higgs field and assign a physically meaningful topology to it? And finally, is the effort worth the result, in other words, is it useful to develop topologies for physical fields and study the systems described by these fields in terms of their corresponding topology? In this paper, we take the first step forward and show that there is an intuitive and physically plausible topology associated with vector and also real, inhomogeneous scalar fields represented as perfect fluids. We also give examples of systems in the study of which the topological approach seems to be useful or at least promising.

The plan of this paper is as follows: in \S\ref{sec:level1}, after a brief review of the fundamental properties of time dependent vector fields, we discuss the recently introduced concept of spatial complexity. A simple time translation operator for vector fields is also introduced, which plays an important role in constructing vector field topology. In \S\ref{stopology}, which introduces the main ideas and results of this paper, we show that the natural metric topology defined using the Euclidean norm is of little interest in physical applications. Instead, a phase space is defined in the context of dynamical systems theory with a built-in topology, which is shown to be the standard topology appropriate for physical applications. Finally, in \S\ref{scalarfields}, we show that certain real, inhomogeneous scalar fields correspond to a vector field whose topology can be defined as the topology of the corresponding scalar field. This is one reason that we mainly focus on vector fields in this paper. In \S\ref{sdiscussion}, we summarize and discuss our results and their physical implications.

\section{\label{sec:level1}Vector Fields}

In this section, we introduce the theoretical tools to be employed in \S\ref{stopology}, where we develop vector field topology. Most importantly, the concept of vector field complexity, eq.(\ref{complexity}) introduced recently in \cite{JV2019}, and the time translation operator, eq.(\ref{TTO}), are briefly discussed.

Let us start with the governing equation of the field. Suppose the real field ${\bf F}$, defined in $n$-dimensional Euclidean space, satisfies a general evolution equation of the following form:

\begin{equation}\label{VFeq}
{\partial\over \partial t}{\bf F}({\bf x}, t)={\bf f}({\bf F}, \partial^\kappa_{\bf x}{\bf F}, {\bf x}, t),
\end{equation}
where the notation $\partial_{\bf x}^\kappa{\bf F}$ is used to imply that $\bf f$ may involve spatial derivatives of order $\kappa\in \mathbb{N}$. For the sake of simplicity, we will use the notation ${\bf f}({\bf x}, t)$ throughout this paper keeping in mind that $\bf f$ may contain $\bf F$ and its spatial derivatives of any order $\kappa\in \mathbb{N}$. In physical problems, the field is often studied in a region of space with some boundaries, thus the problem becomes a boundary value problem which has a unique solution provided that appropriate boundary conditions are applied. We will assume, throughout this paper, the existence of such a unique solution in a spatial volume $V$ without directly referring to any boundary condition. (For a more general definition on manifolds, and also the conditions for the governing equation to have a unique solution, see Appendix \ref{A}.)

How can we quantify the level of spatial entanglement or complexity associated with a vector field? What quantitative measure distinguishes a smooth field, e.g., the velocity field of the laminar flow in a creek, and a spatially complex field, e.g., the velocity field in a turbulent river?

\subsection{Spatial Complexity}\label{sscomplexity}

Before we discuss spatial complexity, let us first illustrate how the governing equation of a given vector field can be decomposed into two differential equations each of which governs either the direction or the magnitude of the field. More details along with some physical implications in electrically conducting fluids can be found in Appendix \ref{spatialcomplexity}.

The unit, direction vector corresponding to $\bf F$, is defined as $\hat{\bf{F}}={\bf F}/F$ while its magnitude is given by $F=|{\bf F}|=\sqrt{F_1^2+...+F_n^2}$, which is of course the Euclidean vector norm. The derivative of the unit, direction vector is given by $\partial_t\hat{\bf{F}}=({\partial_t{\bf{F}}/ F} )_\perp$, and thus eq.(\ref{VFeq}) implies

  \begin{equation}\label{direction}
  \partial_t \hat {\bf{F}}= {{\bf{f}}_\perp\over F},
 \end{equation}
where $(\;)_\perp$ denotes the perpendicular component with respect to $\bf F$. Thus, the direction of $\bf F$ is determined solely by the perpendicular (with respect to $\bf F$) component of $\bf f$. Similarly, it is easy to show that the magnitude of $\bf F$ is determined by the parallel (with respect to $\bf F$) component of $\bf f$. We have

\begin{equation}\label{magnitude}
\partial_t F=\Big(\partial_t {\bf F}\Big)_\parallel={f}_\parallel, \;\;\text{or} \;\; \partial_t\Big(F^2/ 2 \Big)=F{f}_\parallel.
\end{equation}

It follows that 
\begin{equation}\notag
\begin{cases}
\partial_t \hat{\bf F}={\bf 0} \Longleftrightarrow {\bf f}_\perp={\bf 0} \Longleftrightarrow {\bf{f\times F}=0},\\
\partial_t F=0\Longleftrightarrow {\bf f}_\parallel={\bf 0}\Longleftrightarrow {\bf{f. F}}=0.
\end{cases}
\end{equation}
Therefore, pointwise, ${\bf{f\times F}=0}$ and ${\bf{f. F}}=0$ constrain, respectively, the topology and magnitude of the field. 

The spatial complexity of the vector field $\bf F$\footnote{For a more general definition, see eq.(\ref{stochasticity-rate}) in Appendix \ref{spatialcomplexity}.}:

\begin{equation}\label{complexity}
S_2(t)={1\over 2} \Big(  \hat{\bf F}_l.\hat{\bf F}_L -1\Big)_{rms},\end{equation}
where $\hat{\bf F}_l={\bf F}_l/|{\bf F}_l|$ and ${\bf F}_l$ is the average field in a spatial volume of length scale $l$, e.g., defined as

$${\bf F}_l({\bf x}, t)=\int_0^\infty G\Big({{\bf r}\over l}\Big){\bf F}({\bf x+r}, t) {d^3r\over l^3},$$
with $G({\bf r}/l)=G(r/l)$ as a smooth, rapidly decaying function. The large scale field ${\bf F}_L$ with $L\gg l$ is defined similarly. These are renormalized or coarse-grained fields, while $\bf F$ is the bare field. In order to understand the motivation behind the definition (\ref{complexity}), note that $\hat{\bf F}_l.\hat{\bf F}_L=\cos\theta$ is a measure of the angle $\theta$ between the large scale field, ${\bf F}_L$, and small scale field, ${\bf F}_l$, at point $({\bf x}, t)$. A smoothly flowing laminar flow e.g., in a creek, looks almost the same no matter we look at it closely or from a distance. In other words, the velocity field of the water flow is almost the same over all scales, thus $\cos\theta\simeq 1$ and the RMS average $(\cos\theta-1)_{rms}$ would be negligible. A spatially complex vector field, on the other hand, looks very different at different scales, hence $\cos\theta$ would on average deviate from unity and thus $(\cos\theta-1)_{rms}$ would be greater than zero. To have a positive number between $0$ and $+1$, as a measure of the level of spatial complexity of $\bf F$, we can take ${1\over 2} (\cos\theta-1)_{rms}$ which is the same as the definition given by eq.(\ref{complexity}).

As an application of the spatial complexity corresponding to vector fields, defined by eq.(\ref{complexity}), take the magnetic and velocity fields in magnetized fluids as an example. These fields are governed respectively by the induction and Navier-Stokes equations, which like the field themselves, can be coarse-grained. In doing so, it turns out that the evolution of magnetic complexity is closely related to the spatial complexity of the velocity field, which provides a means to study the phenomena of magnetic reconnection and magnetic field generation in astrophysical bodies. For details see \cite{JV2019}; \cite{SecondJVV2019} and \cite{JVV2019}. Appendix \ref{spatialcomplexity} also provides a general review of these recent developments. We will not delve into mathematical details of such applications here, instead our goal of bringing up the concept of complexity in this paper is to compare and relate it to the notion of topological complexity or entropy discussed in \S\ref{ssentropy}.

\subsection{Time Translation and Time Symmetry}

In this subsection, we formulate the time evolution of  vector fields in terms of the time translation operator ${\cal T}(\epsilon)$, which takes the field ${\bf F(x}, t_0)$ at time $t_0$ and maps it into the field ${\bf F(x}, t_0+\epsilon)$ at a later time $t_1=t_0+\epsilon$. The reason we need this concept is as follows: the mathematical operation which deforms a ball to make another topologically equivalent object like a cube is in fact a map. The ball is mapped to the cube: because the deformation does not involve tearing/gluing, it keeps the initial topology of the ball. As mentioned in the Introduction, such a topology-keeping map is called a homeomorphism (onto, one-to-one and continuous maps with continuous inverse). Similarly, as the field ${\bf F(x}, t)$ evolves in time, at any moment $t_0$, the field ${\bf F(x}, t_0)$ is in fact mapped, by the time translation, to the new field ${\bf F(x}, t_0+\epsilon)$ (for an arbitrarily small $\epsilon$). If the old field shares the same topology with the new field, i.e., if the field's topology does not change over time, the time translation ${\cal T}(\epsilon)$ must be a homeomorphism. This will help us, in \S\ref{stopology}, identify the mathematical conditions that should be satisfied by the field and its governing differential equation to keep the field's topology unchanged.

In order to study the time evolution of $\bf F$, assuming ${\bf F}({\bf x}, t_0)$ is given at time $t_0$, we can solve eq.(\ref{VFeq}) to obtain ${\bf F}({\bf x}, t_0+\epsilon)$ at a different time $t_1=t_0+\epsilon$ for an infinitesimal $\epsilon\in \mathbb{R}$ (besides the trivial case of $\epsilon=0$, corresponding to the identity operator, if $\epsilon>0$ we move forward in time, otherwise backward). We can represent this as a linear time translation operator (similar to shift operator in functional analysis or lag operator in time series analysis), $\hat{\cal T}: \mathbb{R}^n\rightarrow \mathbb{R}^n$:

\begin{equation}\label{TTO}
\hat{\cal T}(\epsilon) {\bf F}({\bf x}, t_0)= {\bf F}({\bf x}, t_0+\epsilon),
\end{equation}
which is linear
\begin{eqnarray}\notag
\hat{\cal T}(\epsilon)&& \Big(\alpha{\bf F}({\bf x}_0, t_0)+\beta {\bf F}({\bf x}_1, t_0)  \Big)\\
&=& \alpha{\bf F}({\bf x}_0, t_0+\epsilon)+\beta{\bf F}({\bf x}_1, t_0+\epsilon),\forall \alpha, \beta\in \mathbb{R},\notag
\end{eqnarray}
and its inverse is defined as\footnote{Note that the time translation operator can be written as $
\hat{\cal T}(\epsilon):=e^{\epsilon {\partial\over\partial t} },
$ which is defined operationally in terms of a Taylor series in $\epsilon$. Hence $
e^{\epsilon {\partial\over\partial t} } {\bf F}({\bf x}, t)\equiv{\bf F}({\bf x}, t+\epsilon)$. }

\begin{equation}\label{inverse}
\hat{\cal T}^{-1}(\epsilon)=\hat{\cal T}(-\epsilon),
\end{equation}
with

\begin{equation}\notag
\label{identity}\hat{\cal T}(0)=\hat{\cal I},
\end{equation}

where $\hat{\cal I}$ is the identity operator. Because $\epsilon\in\mathbb{R}$ can be positive or negative, the inverse map, $\hat{\cal T}^{-1}$, is well-defined if the governing equation, eq.(\ref{VFeq}), is invariant under the time reversal operator $\hat\Theta: t\rightarrow -t$;

\begin{equation}\label{VFeq10}
{\partial\over \partial (-t)}{\bf F}({\bf x}, -t)={\bf f}({\bf x}, -t).
\end{equation}
In order for this condition to be satisfied, there are two possibilities: either we have 
\begin{equation}
\begin{cases}
{\bf F}({\bf x}, -t)=+ {\bf F}({\bf x}, t),\\
{\bf f}( {\bf x}, -t)=- {\bf f}( {\bf x}, t),
  \end{cases}
  \end{equation}
  which indicates an even field $\bf F$ and an odd source field $\bf f$, or else we have 
\begin{equation}
\begin{cases}
{\bf F}({\bf x}, -t)=- {\bf F}({\bf x}, t),\\
{\bf f}( {\bf x}, -t)=+ {\bf f}( {\bf x}, t),  \end{cases}
  \end{equation}
 which indicates an odd $\bf F$ and an even $\bf f$. We will see later, in \S\ref{stopology}, that time symmetry plays an important role in the topological formalism of vector fields.

\subsection{Field Lines}
One may wonder if the field lines of a given time-dependent vector field can be used to define a topology for the field. Later, we will use path-lines, which differ from field lines, in the context of dynamical systems and field topology. Therefore, to clarify the different roles of field lines, path lines and also particle trajectories in real space and phase spaces, in this subsection, we briefly review the notion of integral curves or field lines for a given vector field. We also illustrate the fact that, in general, the field lines associated with a given time-dependent vector field cannot be taken as continuously deforming curves in space. Instead, generally speaking, at any moment of time, we may have different field lines.

The field lines of $\bf F$ can be considered as parametric curves whose tangent vector at any point $\bf x$ is parallel to $\bf F$ at that point. At a given time $t_0$, therefore, we can parametrize these curves using the arc-length $s$

\begin{equation} \label{fieldline}
\begin{cases}
{\partial\bxi_{\bx}(s, t_0)\over \partial s} =\hat{\bf F}(\bxi_\bx(s, t_0), t_0),\\
 \bxi_\bx(0, t_0)=\bx,
 \end{cases}
\end{equation} 
where $\hat{\bf F}={\bf F}/|\bf F |$ is the direction (unit) vector. If the unit vector field $\hat{\bf F}$ is Lipschitz continuous\footnote{The real function ${\bf F}(\bx)$ is H{\"o}lder continuous if it satisfies 

$$| {\bf F}({\bx}) - {\bf F}({\bx}') | \leq {\cal F}_0 | {\bx} - {\bx}'|^{\cal H},$$
for ${\cal H}>0$ with some constant ${\cal F}_0>0$. If ${\cal H}=1$, the function is called Lipschitz continuous. If the above condition holds only for $0<{\cal H}<1$, {\bf F} is H{\"o}lder singular. In this case, the derivatives of $\bf F$ are not well-defined in general, thus any function of these derivatives will be generally ill-defined. Also, $\bf F$ is bi-Lipschitz if for some $F_0>0$;
$${1\over{F}_0} | {\bx} - {\bx}'|\leq | {\bf F}({\bx}) - {\bf F}({\bx}') | \leq{F}_0 | {\bx} - {\bx}'|.$$
}in $\bx$, then the above differential equation has a unique solution\footnote{In general, for ${\bf F}: \mathbb{R}^n\rightarrow \mathbb{R}^n$ restricted onto a curve $\boldsymbol\xi(s)$ parametrized by $s\in \mathbb{I}\subseteq \mathbb{R}$, i.e., ${\bf F}={\bf F}(\boldsymbol\xi(s))$, suppose each $F_j$, for $1\leq j\leq m$, is continuous on $\mathbb{I}\times \mathbb{R}^n$ and additionally is uniformly Lipschitz continuous, i.e., $| F_j(\boldsymbol\xi)-F_j(\boldsymbol\eta)  |\leq {\cal{F}}_0 |\boldsymbol\xi-\boldsymbol\eta|$ for some ${\cal{F}}_0>0$. If for $s_0\in \mathbb{I}$, there are real numbers $c_1,...c_n$ such that ${\boldsymbol\xi}_j(s_0)=c_j$, then there exists a unique solution $\boldsymbol\xi(s)$ for the differential equation $d\boldsymbol\xi/ds={\bf F}(\boldsymbol\xi(s))$ with initial conditions $\boldsymbol\xi_j(s_0)=c_j$. This is the Picard-Lindel{\"o}f uniqueness theorem for a system of differential equations. For the simpler problem $y'(s)=f(s,y(s))$ with $ y(s_0)=y_0$, the theorem indicates that if $f$ is uniformly Lipschitz continuous in $y$, and continuous in $s$, then for some $\delta>0$, a unique solution exists on the interval $[s_0-\delta, s_0+\delta]$.} and hence there exists a unique field-line $\bxi_\bx(s, t_0)$ passing through $\bx$ at time $t_0$. Nevertheless, if the field is H\"older singular, then there may exist infinitely many such integral curves (solutions) satisfying the above differential equation\footnote{The Peano theorem can still be used here to infer the existence of solutions, however, the uniqueness of a solution is guaranteed by the Picard-Lindel{\"o}f theorem which requires Lipschitz continuity of $\bf F$ as discussed before.}. Thus for H{\"o}lder-singular fields, the concept of field line is not generally well-defined. On the other hand, for a renormalized field ${\bf F}_l$, integration by parts shows that any spatial derivative $\nabla_x$ acting on ${\bf F}_l$ can be made to act on $G$ instead of $\bf F$ inside the integral, which implies that ${\bf F}_l$ is Lipschitz-continuous even if ${\bf F}$ is not. Thus the notion of field line is well-defined for the renormalized field ${\bf F}_l$ even if ${\bf F}$ is H{\"o}lder singular.

In order to understand how the field lines evolve in time, we write the defining equation of the integral curves, given by eq.(\ref{fieldline}), at another time $t_0+\epsilon$ for a real $\epsilon$ as

 \begin{equation} \label{fieldline4}
\begin{cases}
{\partial\bxi_{\bx}(s, t_0+\epsilon)\over \partial s} =\hat{\bf F}(\bxi_\bx(s, t_0+\epsilon), t_0+\epsilon),\\
 \bxi_\bx(0, t_0+\epsilon)=\bx. 
 \end{cases}
\end{equation} 

The condition for $\bxi_\bx$ to be continuous in $t$ is $\lim_{\epsilon\rightarrow 0}|\bxi_\bx(s, t_0+\epsilon)-\bxi_\bx(s, t_0)|\rightarrow 0$ for arbitrary curve parameter $s$. This requires (Appendix \ref{fieldlinecontinuity})

\begin{equation}\label{continuity}
\Big| \bxi_{\bx}(s, t_0+\epsilon)-\bxi_{\bx}(s, t_0)   \Big|\leq |\epsilon| \Big( {e^{K_0 s}-1\over K_0} \Big),
\end{equation}
for some $K_0>0$. For any finite but arbitrarily large $s>0$, we can take $|\epsilon|$ small enough to make the RHS of (\ref{continuity}) arbitrarily small, which indicates that ${\bxi}_\bx$ is uniformly continuous in time. Consequently, $\bxi_\bx(s, t)$ is uniformly continuous in $t$, provided that $\hat{\bf F}({\bx}, t)$ is Lipschitz\footnote{Note that one may also use the Minkowski metric here, which is
$$|\vec{x}_2-\vec{x}_2|=\sqrt{ |{\bf x}_2-{\bf x}_1|^2 -|t_2-t_1|^2      }.$$
In any case, the continuity of the integral curves in time requires continuity of $\bf F$ in spacetime and not just space.}  in $\vec{x}=({\bx}, t)$. Lipschitz continuity of $\hat{\bf F}$ in $\vec{x}=({\bx}, t)$ indicates that $\hat{\bf F}$ is Lipschitz in both $\bx$ and $t$ (which can be seen from the last line of eq.(\ref{continuity1}) in Appendix \ref{fieldlinecontinuity}). 

As an application of the concepts discussed in this subsection, consider magnetic field lines in a dissipative and turbulent fluid, i.e., a real fluid with finite electrical resistivity and viscosity such as most of astrophysical plasmas. Magnetic fields in such environments are usually visualized in terms of their field lines, which may undergo a spontaneous process of breaking apart and re-connecting again, which may accelerate the magnetized fluid by means of the Lorentz forces (i.e., magnetic reconnection; for a review see e.g., \cite{JafariandVishniac2018} and \cite{Review2020}). However, as discussed above, the magnetic field lines in such  dissipative and dynamic media cannot be described as continuously evolving curves in space. Although the field lines may be well-defined at any moment of time, they have no identity preserved in time and, at any later time, we will have other distinct field lines. The other complication arises due to  the stochastic nature of field lines in real turbulent fluids e.g., in astrophysics. Only recently have these important facts taken into account in problems related to magnetic field generation, evolution and reconnection (see e.g., \cite{Eyink2011}; \cite{Eyinketal2013}; \cite{Eyink2015}; \cite{JV2019}; \cite{JVV2019}). The mere purpose of briefly discussing these concepts here is to emphasize that the notion of field lines cannot be used in a rigorous manner to define or even visualize the topology of a time-dependent vector field. We will see in the next section that path-lines may provide a more useful tool as far as the field topology is concerned.

\section{Topology}\label{stopology}

In this section, we present the major result of this paper using the tools introduced in the preceding sections---the metric topology associated with a given vector field ${\bf F(x}, t)$. 

Let us start with a general remark on the notion of topology in mathematics. Topology is a mathematical concept used to study the properties of an object which are preserved under stretching, bending or twisting but not tearing or gluing. An play dough can be easily deformed, without tearing, to make infinitely many different shapes. But all these shapes are topologically equivalent. However, if we tear the play dough while deforming, we will get separate, disconnected objects. What distinguishes deforming without tearing from deforming with tearing/gluing is that, in the former case, arbitrarily close points remain arbitrarily close while, in the latter case, they do not. One way of defining a topology for an object, therefore, is to use the notion of distance (although it can be defined in other ways as well). A donut, i.e., a torus, made of play dough can be deformed, without tearing or gluing, to make a coffee mug. Thus a donut is topologically equivalent to a coffee mug, however, it is not equivalent to a ball because it cannot be deformed to make a ball without tearing or gluing. That the donut is topologically equivalent to the coffee mug implies that any pair of arbitrarily close points on the donut, say $\bf x$ and $\bf y$, will be mapped into arbitrarily close points on the mug. Mathematically, we can say that two nearby points $\bf x$ and $\bf y$ on the donut, i.e., points with an arbitrarily  small distance $|\bf x-\by|$, are mapped to two arbitrarily close points $\bf x'$ and $\bf y'$ on the coffee mug, i.e., points with an arbitrarily small distance $|\bf x'-\by'|$. 

One of the definitions of a topological space, in terms of open sets, is as follows: take an arbitrary non-empty set $X$ and a collection of its subsets, $\tau$, which are called open sets. The pair $(X, \tau)$ is a topological space if (i) the empty set and $X$ both belong to $\tau$, (ii) any arbitrary union of members of $\tau$ is in $\tau$, and (iii) the intersection of any finite number of subsets of the topology $\tau$ belongs to $\tau$. One can equivalently use closed sets instead, in which case (i) the empty set and $X$ both belong to $\tau$, (ii) any finite union of members of $\tau$ is in $\tau$, and (iii) the arbitrary intersection of of subsets of the topology $\tau$ belongs to $\tau$. As an example, the familiar Euclidean space $\mathbb{R}^3$ is a topological space. This can be easily verified using the the definition of Euclidean distance between any two points ${\bf x, y}\in \mathbb{R}^3$. This distance, called metric, helps us define an open ball $B_r({\bf x}_0)$ around any arbitrary point ${\bf x}_0\in \mathbb{R}^3$ as the set of all points whose distance from ${\bf x}_0$ is less than a positive number $r>0$:
$$B_r({\bf x}_0)=\{{\bf x}\in \mathbb{R}^3\mid \;|{\bf x}-{\bf x}_0|<r      \},$$
which is called an open ball of radius $r>0$ around ${\bf x}_0$. These balls form a base for the topology on $ \mathbb{R}^3$ because any open set of $ \mathbb{R}^3$ can be given by an arbitrary union of these open balls. 

Of course, we could consider more general spaces, e.g., $n$-dimensional space $\mathbb{R}^n$, $n\in\mathbb{N}$, which is also a topological space. Generally speaking, any set ${M}$ equipped with a metric\footnote{A metric $d$ on a set $M$ is a function $d: M\times M\rightarrow R$ such that for any $x, y, z\in M$ such that $(i)\;d(x, y)=0\Leftrightarrow x=y$, $(ii)\; d(x, y)=d(y, x)$, $(iii)\; d(x, z)\leq d(x, y)+d(y, z).$}, i.e., an appropriate notion of distance $d( x, y)$ between any pair of its members $x, y\in {M}$, is called a metric space. For instance, $ \mathbb{R}^n$ is a metric space with the metric $d({\bf x}, {\bf y})=|{\bf x}-{\bf y}|$ as discussed above. Any metric space is a topological space, because we can use the metric to define open balls as the base for the topology as we did for $ \mathbb{R}^3$.

\subsection{Vector Filed Topology}

As discussed previously, any set equipped with a metric, i.e., a notion of distance between its elements, is a metric space, and any metric space is a topological space. On the other hand, any vector field $\bf F({\bx})$ comes with a natural metric induced by the vector norm, i.e., the distance between $\bf F({\bx})$ and $\bf F({\by})$ can be defined as $|{\bf F}({\bx})-\bf F({\by})|$. Hence, a vector field as a set of vectors in ${\mathbb R}^n$ is a topological space. In order to construct a topology, for a given vector field $\bf F$, let us define an open ball ${\cal B}_{\cal R}({\bf F}({\bx}_0, t_0))$, with radius ${\cal R}>0$, around the vector ${\bf F}({\bx}_0, t_0)$ as 

\begin{equation}\label{openball1}
{\cal B}_{\cal R}({\bf F}({\bx}_0, t_0))=\{ {\bf F}({\bx}, t_0) \mid \;      | {\bf F}({\bx}, t_0)-{\bf F}({\bx}_0, t_0)   |<{\cal R}\}.
\end{equation}

It should be emphasized that, in general, this ball is not localized in real Euclidean space ${\mathbb R}^n$, i.e., it doesn't imply $|{\bf x}-{\bx}_0|<c$  for some $c>0$, unless $\bf F$ is biLipschitz, that is to say there is an $F_0>0$ such that ${1\over F_0} |{\bf x}-{\bx}_0|\leq | {\bf F}({\bx}, t_0)-{\bf F}({\bx}_0, t_0)   |\leq F_0 |{\bf x}-{\bx}_0| $. Therefore, for a general, non biLipscitz field, the ball defined by eq.(\ref{openball1}) is a set of vectors ${\bf F}({\bx}, t_0)$ which are close to ${\bf F}({\bx}, t_0)$ as measured by the metric induced by the vector norm. If we visualize the field in space, e.g., a three-dimensional magnetic field around a magnet, then we will see that the ball defined above is a set of vectors which are scattered in space. If we define open sets using these open balls as the bases, we will have a metric topology for the vector field $\bf F$.

Let us consider the conditions under which the field $\bf F$ keeps its topology as it evolves in time. In other words, we look for conditions under which the time translation operator $\hat{\cal T}(\epsilon)$ acting on a given vector field $\bf F$ is a homeomorphism, i.e., a continuous, bijective map with continuous inverse. Continuous topological deformations do not change the topology; one can deform (i.e., map) an object such as a donut (i.e., a topological space) without cutting and gluing it (i.e., a continuous map which takes nearby points to nearby points) to a coffee mug (i.e., another topological space). Conversely, one should be able to recover the donut by deforming the coffee mug (i.e., the continuous map should have a continuous inverse). In addition, all points on the donut should be mapped; each point to only one point (i.e., an onto and one-to-one map). Instead of two topological spaces $X_0$ and $X_1$, e.g., a donut and a coffee mug, one may consider a vector field at two different times $t_0$ and $t_1=t_0+\epsilon$. With our definition for vector field topology given above, the time translation operator $\hat{\cal T}$ should be a homeomorphism to preserve the topological properties of the field as it evolves in time.

In order for $\hat{\cal{T}}(\epsilon)$ to be one-to-one, it should satisfy the following condition:

$$\hat{{\cal T}}(\epsilon) {\bf F}({\bx}_1, t_0)=\hat{{\cal T}}(\epsilon) {\bf F}({\bx}_2, t_0)\rightarrow {\bf F}({\bx}_1, t_0)= {\bf F}({\bx}_2, t_0),$$

that is
\begin{equation}\notag
{\bf F}({\bx}_1, t_0+\epsilon)={\bf F}({\bx}_2, t_0+\epsilon)\rightarrow {\bf F}({\bx}_1, t_0)= {\bf F}({\bx}_2, t_0).
\end{equation}

Obviously, this condition will not be satisfied in general. In other words, even for smooth and well-defined vector fields, the time evolution operator $\hat{\cal T}$ is not in general a homeomorphism, hence the natural topology of the vector field can change over time no matter how smooth the field $\bf F$ is. The other issue is that the open balls, i.e., the bases defining open sets for the field's metric topology, are not necessarily (except for biLipschitz fields) localized in real space as discussed above. This is extremely restricting in physical applications. It is more desirable to define a topology such that, roughly speaking, open balls are associated with nearby points in both real space and vector space. Such an open ball around ${\bf F}({\bx}_0, t_0)$, at a given time $t_0$, is defined as (with a proper non-dimensionalization)

\begin{eqnarray}\notag
&&{\cal B}_r({\bf F}({\bx}_0, t_0))=\{ ({\bf F}({\bx}, t_0)) \mid \; \\
&&\sqrt{ |{\bx}-{\bx}_0|^2+|{\bf F}({\bx},t_0)-{\bf F}({\bx}_0, t_0)|^2}<r\}.
\end{eqnarray}
In fact, this corresponds to an open ball in the phase space $({\bx}, {\bf F})\equiv (x, y, z, F_x, F_y, F_z)$. These open balls, therefore, define a topology in this phase space. In the next subsection, we will briefly study this phase space and its topology.

\subsection{Phase Space and Dynamical Systems}

In the previous subsection, we argued that a physically plausible notion of topology for a vector field $\bf F$ can be defined as the metric topology of the phase space $(\bf x, F)$. The corresponding dynamical system is defined as\footnote{A dynamical system is a system whose time evolution is described by an initial value problem. For example, a planet orbiting around the sun has a velocity ${\bf u}=d{\bf x}/dt$, governed by the Newton's second law, $d{\bf u}/dt={\bf F}$ where $\bf F$ is the gravitational force. Depending on the initial position of the planet ${\bf x}(t_0)={\bf x}_0$, its future orbit is determined. Classically, the state of the planet is given by its instant position and velocity (momentum). Hence, the space of all possible configurations is $({\bf x, u})$ which is called the phase space. The phase space is defined similarly in classical statistical mechanics as the space of all possible coordinates $q$ and momenta $p$: $(q, p)$. Quantum mechanics can also be expressed in phase space as a deformation of classical mechanics with deformation parameter $\hbar/\text{(action)}$ similar to the deformation of Newtonian mechanics into special relativistic mechanics with deformation parameter $v/c$.} 

\begin{equation}\label{DSeqs}
{d{\bx}\over dt}={\bf F}, \;\;\;{\partial{\bf F}\over \partial t}={\bf f},
\end{equation}
which define a dynamical system. Assuming that the second equation in (\ref{DSeqs}) uniquely determines $\bf F$ (e.g., with appropriate boundary conditions), the first equation requires an initial condition in the form 

$${\bx}(t_0)={\bx}_0$$

 to have a unique solution as a trajectory in the phase space (with uniqueness guaranteed by applying appropriate conditions on $\bf F$ e.g., Lipschitz continuity). However, we are not interested in any particular trajectory here; we are interested in all possible trajectories ${\bf x}={\bf x}(t)$ that the field $\bf F$ defines in the phase space $(\bf x, F)$. Each trajectory is determined by its corresponding initial condition and can be visualized as the trajectory (in the phase space) of a particle moving with the time dependent velocity ${\bf F}$. The trajectories are solutions of the following non-autonomous\footnote{The differential equation $d{\bx}(t)/dt={\bf F}({\bx}(t), t)$ is a non-autonomous equation because $\bf F$ explicitly depends on time $t$. Dynamical systems governed by such non-autonomous equations are much more difficult to study. One might naively attempts to eliminate this time-dependence by introducing a new variable, $\tau(t)$ such that $d\tau(t)/dt=1$. With this choice, the equation $d{\bx}(t)/dt={\bf F}({\bx}(t), \tau(t))$ becomes autonomous. However, this does not simplify the task of solving the equations since the dimension of the problem is increased by introducing a new function. }  differential equation (cf. (\ref{fieldline}) in the previous section):
 
 \begin{equation}\label{DSeqs2}
\begin{cases}
{d{\bx}(t)\over dt}={\bf F}({\bx}(t), t),\\
{\bx}(t_0)={\bx}_0,
\end{cases}
\end{equation}
 
which has a unique solution if $\bf F$ is uniformly Lipschitz continuous in $\bx$ and continuous in $t$ (cf. Lipschitz continuity of $\hat{\bf F}$ in $\bx$ for eq.(\ref{fieldline}) to define unique field lines). The time translation operator, acting at any point $({\bx}, {\bf F}({\bx}, t))$ in the phase space, can be represented as

\begin{equation}
\hat{\cal T}_e(\epsilon) ({\bx}, {\bf F}({\bx}, t))=({\bx}, {\bf F}({\bx}, t+\epsilon)).
\end{equation}

It is easy to see that $\hat{\cal T}_e(\epsilon)$ is an onto, one-to-one, and continuous map with continuous inverse. For its continuity, for example, we note that $\hat{\cal{T}}_e(\epsilon)$, for any $\epsilon \in \mathbb{R}$, is continuous (so is its inverse for $\hat{\cal{T}}_e^{-1}(\epsilon)=\hat{\cal{T}}_e(-\epsilon)$) if it is continuous at $\epsilon=0$. In order to show this for any $t$, the following ${\cal L}_1$-norm should vanish in the limit $\epsilon\rightarrow 0$,

\begin{equation}\notag
\lim_{\epsilon\rightarrow 0}  \int_{t\in {\mathbb{I}}_t} dt \Big| \hat{\cal{T}}_e(\epsilon)({\bx}, {\bf F}({\bx}, t))-\hat{\cal{T}}_e(0)({\bx}, {\bf F}({\bx}, t) )\Big|.
\end{equation}
Thus the condition for the continuity of $\hat{\cal{T}}_e(\epsilon)$ is
\begin{equation}
\lim_{\epsilon\rightarrow 0}  \int_{t\in {\mathbb{I}}_t}  dt \Big|{\bf F}({\bx}, t+\epsilon)-{\bf F}({\bx}, t) \Big|\rightarrow 0,
\end{equation}

which follows if $\bf F$ is uniformly continuous in $t$.\footnote{Similarly, the shift operator defined by $\hat{\cal{O}}_af(x)=f(x+a)$ is continuous if $f$ has compact support and is continuous, which implies that $f$ is uniformly continuous.} In order to keep the phase space topology preserved in time, we need to ensure that the phase space at any given time $t_0$, as a topological space,, is homeomorphic to the phase space at another time $t_1$. The condition of continuity for $\hat{\cal{T}}_e^{-1}(\epsilon)=\hat{\cal{T}}_e(-\epsilon)$, on the other hand, requires the equations given by (\ref{DSeqs}) to be time reversal invariant, which requires $\bf F$ to be odd, i.e., ${\bf F}({\bx}, -t)=-{\bf F}({\bx}, t)$ and $\bf f$ to be even, i.e., ${\bf f}({\bx}, -t)=+{\bf f}({\bx}, t)$.

In short, the topology associated with a given vector field $\bf F$ is naturally defined in the phase space $(\bf x, F)$. Moreover, in order to ensure that time evolution keeps the topological properties of the phase space, the field ${\bf F}({\bx}, t)$ is required to be (i) Lipschitz continuous in $\bx$, (ii) uniformly continuous in $t$ and (iii) odd under time reversal, i.e., ${\bf F}({\bx}, -t)=-{\bf F}({\bx}, t)$, such that its governing equation, $\partial_t{\bf F}={\bf f}$, eq.(\ref{VFeq}), is time reversal invariant.

As for physical applications, once again, we can think about magnetic fields. Magnetic topology in magnetized fluids is often claimed to change during magnetic reconnection---eruptive fluid motions spontaneously driven by Lorentz forces generated by the relaxing field, see e.g., \cite{Yamadaetal2010}; \cite{Review2020}. Magnetic field is governed by the induction equation and the velocity field by Navier-Stokes equation. Both turbulent and small-scale plasma effects appear in these differential equations as non-ideal terms which break the time symmetry. A detailed consideration of magnetic and velocity field topologies in such systems is beyond the scope of this paper and will be presented elsewhere, however, here we note that this topological approach relates phenomena such as magnetic reconnection to more fundamental concepts such as time reversal invariance. The presence of non-ideal terms in the field equations break the time symmetry, which is related to entropy increase, thus violating one of the conditions to have a preserved topology. In fact, the presence of dissipative phenomena and also turbulence will in general indicate that the velocity and magnetic fields are also H{\"o}lder singular \cite{JV2019}, therefore, real magnetic fields e.g., on the solar surface, are expected to change their topology frequently.

\subsection{Topological Entropy}\label{ssentropy}

In \S\ref{sscomplexity}, we introduced a measure for the spatial complexity of a vector field. On the other hand, as discussed in the previous subsection, a well-defined time dependent vector field defines a phase space and a dynamical system. The complexity of dynamical systems is quantified using a mathematical concept called topological entropy. Therefore, it is interesting to see if the vector field complexity has anything to do with the topological complexity associated with the dynamical system defined by that vector field. In this subsection, we consider this question, although in a very brief form as a more detailed consideration is obviously out of the scope of this paper.

In a given phase space, we are interested to distinguish two distinct groups of points in a neighborhood of a given trajectory: those points whose distance grows over time as the system evolves from those points whose distance does not. This can be made precise in terms of a metric \cite{Bowen1971}, although there are other equivalent ways to do so, e.g., in terms of covers in compact Hausdorff spaces \cite{Adleretal1965}. The topological entropy is a way of counting the number of distinct trajectories which are generated as the dynamical system evolves in time\footnote{For a compact metric space $(M, d)$ equipped with a continuous map $g: M\rightarrow M$, one can define for each $n\in \mathbb{N}$, the metric (\cite{Bowen1971}), $d_n: M\times M\rightarrow {\mathbb{R}}$ as

$$d_n(x,y)=\max\{d(g^k(x),g^k(y)): 0\leq k \leq n-1\},$$

for any $x, y\in M$. For any real, positive $\epsilon$ and $n\geq 1$, two points of $M$ are said to be $\epsilon$-close if their first $n$ iterates are $\epsilon$-close. A subset $N\subseteq M$ is $(n, \epsilon)$-separated provided that the distance between every distinct points of the subset $N$ is larger than or equal to $\epsilon$ as measured by the above metric. Suppose $\cal N$ is the maximum cardinality of such an $(n, \epsilon)$-separated set, which is a finite number because of the compactness of $M$. The topological entropy of the map $g$, as a measure of the complexity of the corresponding dynamical system, is a non-negative, real number defined as

$$h(g)=\lim_{\epsilon\to 0} \left(\limsup_{n\to \infty}\frac{1}{n}\log {\cal N}(n,\epsilon)\right).$$}. For a dynamical system governed by a given iterated function, the topological entropy can be thought of as a measure of the exponential growth rate of the number of distinguishable orbits, which is an extended real number. In other words, topological entropy is a measure of the system's complexity level. 

One can also consider the vector field $d{\bx}/dt=\bf F$ in the phase space $({\bx}, {\bf F})$ and ask how the complexity of the dynamical system, measured by the entropy, manifests itself in terms of the vector field $\bf F$. The field ${\bf F}({\bx}, t)$, as a function of $\bx$, is the vector tangent to the trajectory ${\bx}(t)$. As the system evolves, therefore, the vector field $\bf F$ determines the direction of motion in the phase space. Consequently, two distinct points on two close trajectories in the phase space will remain close if the corresponding tangent vectors of their trajectories remain close (in the tangent bundle). This argument suggests that the topological entropy is intimately related to the evolution of the spatial configuration associated with the corresponding vector field, i.e., the spatial complexity defined by eq.(\ref{complexity}). For example, the velocity field corresponding to a laminar flow retains its untangled and smooth configuration in time, unlike the entangled and complex velocity field corresponding to a fully turbulent flow. These are expected to be associated respectively with lower and higher entropies in the corresponding phases spaces.

 \section{Scalar Fields}\label{scalarfields}
 
There is a certain type of scalar fields which can be 
presented as a perfect fluid with a well-defined velocity field in $\mathbb{R}^3$. The scalar field topology can then be studied in terms of this vector field.
The main purpose of this final section, therefore, is to illustrate how an appropriate vector field can be defined for an inhomogeneous (i.e., a field with dependence on spatial coordinates $\bf x$) scalar field $\phi({\bf x}, t)$. Examples include the real, inhomogeneous inflaton field (believed, in inflationary cosmology, to  cause an exponential expansion before eventually decaying to baryonic matter/radiation in the early universe) as well as quintessence (the field corresponding to a time-dependent vacuum energy) and Higgs field (the field whose coupling to matter fields gives particles their mass).

Consider the Lagrangian
  
\begin{equation}\label{lagrangian1}
{\cal L}=-\Big( {1\over 2} g_{\mu\nu}\partial^\mu \phi \partial^\nu \phi +V(\phi) \Big),
\end{equation}
with an arbitrary potential function $V(\phi)$. The action $S=\int d^4x \sqrt{-\text {det} g_{\mu\nu}  }{\cal L}(\phi, \partial_\mu \phi)$  is by definition stationary with respect to general coordinate transformations, therefore, we get the corresponding energy-momentum tensor as $T^{\mu\nu}=g^{\mu\nu}{\cal L}+\partial^\mu\phi\partial^\nu\phi$. Comparing this to the energy-momentum tensor of a perfect fluid with density $\rho$, pressure $P$ and four-velocity $u^\mu={dt\over d\tau}(1, {\bf v})$, i.e., $T^{\mu\nu}=Pg^{\mu\nu}+(P+\rho)u^\mu u^\nu$, it is straightforward to show (Appendix \ref{Vectors}) that the perfect fluid velocity is given by 
 
\begin{equation}\label{fluidvelocity}
{\bf v}({\bf x}, t)=-{\nabla\phi({\bf x}, t)\over \dot\phi({\bf x}, t)},
\end{equation}
  
provided that $\dot\phi\equiv \partial_t\phi\neq 0$.  This velocity, eq.(\ref{fluidvelocity}), solves the relativistic Euler equation:  
  \begin{eqnarray}\label{Euler10}
{\partial {\bf v}\over \partial t}+{\bf v}.\nabla{\bf v}&=&-{1-v^2\over P+\rho}\Big( \nabla P+{\bf v}{\partial P\over \partial t} \Big),
\end{eqnarray}
where $v\equiv |\bf v|$. The fluid pressure is given by $P=-{1\over 2}g_{\mu\nu}\partial^\mu\phi\partial^\nu\phi-V(\phi)$ and the density by $\rho=-{1\over 2}g_{\mu\nu}\partial^\mu\phi\partial^\nu\phi+V(\phi)$. 

A non-ideality associated with $\phi$, in this picture, would add terms to the Lagrangian, the energy-momentum tensor and consequently the Euler equation above. For example, in inflationary cosmology, the inflaton field has to eventually couple to normal matter/radiation in order to decay and give rise to the standard radiation dominated Big Bang cosmology in the early universe. This is usually accounted for by adding a decay term to the Lagrangian, which modifies the equation of motion. Consequently, with additional terms in the energy-momentum tensor, the fluid becomes imperfect whose Euler equation contains additional non-ideal terms. Such terms in turn can break the conditions required for an invariant topology, such as time reversal invariance. Thus the inflaton decay may be identified with a topology change. 

Similar arguments can possibly be advanced for other fields with interesting physical implications, however, we will not pursue these considerations here. This section should in fact be taken only as a short remark to remind us that the scalar fields, that come with a perfect fluid representation, define a natural velocity field $\bf v$ with a metric topology in the corresponding phase space $\bf(x, v)$.

\section{Summary and Conclusions}\label{sdiscussion}

Vector fields in physics are usually visualized in terms of their integral curves, or field lines, while the field topology is implicitly used, in fact, only as a synonym for the field configuration in terms of its field lines. This vague picture differs from the mathematical notion of topology in the context of topological spaces. In fact, in order to have well-defined field lines for a given vector field $\bf F$, the Lipschitz-continuity of $\hat{\bf F}={\bf F}/F$ in spatial position vector $\bx$ is required. In order to have continuously deforming field lines in time, the stronger condition of the Lipschitz continuity of $\hat{\bf F}$ in spacetime position vector $\vec{x}=({\bx}, t)$ should also be satisfied. These requirements are not met in many applications, e.g., velocity and magnetic fields in turbulent plasmas are non-Lipschitz and also have stochastic behavior. 
As for the field's topology and topology change, these notions are usually employed in physical applications, e.g., in plasma physics literature in the context of magnetic fields threading highly conducting fluids, to indicate a spontaneous change in the field configuration in terms of its field lines. This is once again different from what a topology change means in mathematics, i.e., maps between topological spaces which fail to be a homeomorphism and therefore may change the topological properties. 

In order to study the  topology change of a given vector field $\bf F$, one needs to (i) define vector field topology using the standard mathematical formalism e.g., in terms of open sets, (ii) define a time translation map which takes a time dependent vector field as a topological space at time $t_0$ and maps it to another vector field at another arbitrary time $t_1$ and finally (iii) find the conditions under which such a time translation map is a homeomorphism, i.e., a continuous, bijective map with continuous inverse (for only homeomorphisms, by definition, preserve topological properties). Scalar fields, such as the inhomogeneous inflaton and quintessence fields, define a perfect fluid velocity field, which can be used as a vector field to define the scalar field topology.

The Euclidean vector norm defines a metric for the vector field $\bf F$, as $d({\bf F}({\bx}), {\bf F}({\by}))=| {\bf F}({\bx})-{\bf F}({\by})  |$. With this metric, the vector field defines a metric space and hence a topological space since there is a natural notion of distance between any pair of vectors ${\bf F}({\bx})$ and ${\bf F}({\by})$. However, with this definition, the time translation of $\bf F$ is not a homeomorphism in general, thus this topology is not preserved and hence not interesting physically. In other words, the topology defined by ${\bf F( x}, t_0)$ may differ in general from that defined by ${\bf F( x}, t_1)$ for $t_0\neq t_1$. Furthermore, in physical applications, one is interested in vectors which are not only close in the above sense but also are located at nearby points in real space. For example, when the magnetic field threading a plasma undergoes reconnection, we are concerned with the magnetic vectors in a spatial volume, i.e., the reconnection region. Thus we are interested in the vectors ${\bf F}({\bx})$ and ${\bf F}({\by})$ for which

$$\sqrt{ |{\bx}-{\by}|^2+|{\bf F}({\bx})-{\bf F}({\by})|^2}<r$$

for some $r>0$, which naturally defines an open ball in the phase space $({\bx}, {\bf F})$. Any trajectory $\bx (t)$ in this phase space is a solution to an initial value problem $d{\bx}(t)/dt={\bf F}({\bx}(t), t)$, ${\bx}(0)={\bx}_0$. Therefore, the vector field topology is naturally defined as the phase space topology. In order to ensure that time evolution preserves the topological properties of the phase space, the field ${\bf F}({\bx}, t)$ is required to be (i) Lipschitz continuous in $\bx$, (ii) uniformly continuous in $t$ and (iii) odd under time reversal ${\bf F}({\bx}, -t)=-{\bf F}({\bx}, t)$ with a time reversal invariant governing equation $\partial_t {\bf F}=\bf f$.

Field lines, which are defined only for Lipschitz fields and do not continuously transform in time unless strong conditions are satisfied, are not very appropriate objects in the study of the field topology, time evolution and topology change. Instead the trajectories in the phase space $({\bx}, {\bf F})$ provide better means for such considerations. Dynamics and statistics of physical vector fields, such as stochastic and H{\"o}lder singular magnetic and velocity fields in turbulent plasmas can be studied in the context of mathematical theory of non-autonomous dynamical systems. One implication is that we can also take the phase space trajectories as curves in real Euclidean space. This translates into the fact that instead of field lines (or streamlines), defined by

$$\forall t_0\in \mathbb{R}:\;{\partial\bxi_{\bx}(s, t_0)\over \partial s} =\hat{\bf F}(\bxi_\bx(s, t_0), t_0),\;\text{with}\; \bxi_\bx(0, t_0)=\bx,$$

one can use path-lines in real space defined by

$${d{\bx}(t)\over dt}={\bf F}({\bx}(t), t) \;\text{with}\;\; {\bx}(0)={\bx}_0.$$
This may provide a more appropriate presentation of vector fields in physical applications.

\appendix

\section{General Definition of Vector Fields}\label{A}

In general, a vector field $\bf F$ can be defined as a map from a manifold $\mathbb{M}$ to its tangent bundle $\mathbb{TM}$:

\begin{eqnarray}\notag
{\bf F}: \;&&\mathbb{M} \rightarrow \mathbb{TM},\\\notag
&&\;{\bf x}\rightarrow {\bf F}({\bf x}),
\end{eqnarray}

such that the image of ${\bx}\in\mathbb{M}$, i.e., ${\bf F}({\bx})$, lies in the tangent space at $\bx$, i.e., $\mathbb{T}_\bx\mathbb{M}$. In this paper, we will assume $\mathbb{M}=\mathbb{R}^n$, i.e., the real $n$-dimensional Euclidean space, unless stated otherwise. We will consider a real, time-dependent vector field ${\bf F}({\bx}, t)$ such that
 
 $${\bf F}: \mathbb{R}^{n}\times  \mathbb{I}_t\rightarrow \mathbb{R}^n,$$
 
where $t\in \mathbb{I}_t\subseteq \mathbb{R}$, with the notation ${\bf F}({\bf x}, t)={\bf F}(\vec x)=(F_1(\vec x),..., F_n(\vec x))$, where $\vec x=({\bf x}, t)=(x_1,..., x_n; t)$. If the field is defined for all times, i.e., $\mathbb{I}_t=\mathbb{R}$, we have a flow, otherwise a semi-flow. It goes without saying that a straightforward generalization of multivariate calculus can be applied to vector fields, e.g., $\bf F(x)$ is continuous if $\lim_{\bf x\rightarrow a} {\bf F(x})= {\bf F(a})$.

A general governing equation can take the following form in terms of the field components $F_j$;
\begin{equation}\notag
{\partial^{n_i} F_i\over \partial t^{n_i}}=f_i(t, x_1,...,x_n, F_1,...F_n,..., {\partial^k F_j\over \partial t^{k_0}\partial x_1^{k_1}...\partial x_n^{k_n}},...),
\end{equation}
with $i, j=1,2,...,n$; $k=k_0+k_1+...+k_n\leq n_j$; $k_0<n_j$. The initial conditions are given at time $t=t^0$ on a hyper-surface (called Cauchy data) in the following form:
\begin{equation}\notag
{\partial^k \over \partial t^k}F_i(t^0, x_1,..., x_n)=g_i^{(k)}(x_1,..., x_n), 
\end{equation}
with $k=0, 1, 2,..., n_i-1$. The Cauchy-Kowalevski theorem guarantees a unique solution for the above initial value problem providing that the functions $f_j$ and $g_j$ are analytic. More precisely, if all $f_j$ functions are analytic in a neighborhood of the point $(t^0, x_1^0,..., x_n^0, F^0_1,..., F^0_n, ..., {\partial^k F_j\over \partial t^{k_0}\partial x_1^{k_1}...\partial x_n^{k_n}}|_0,...)$
and all functions $g_j^{(k)}$ are analytic in some neighborhood of $(x_1^0,..., x_n^0)$, then the above Cauchy problem has a unique, analytic solution in some neighborhood of $(t^0, x_1^0,..., x_n^0)$.

\section{Physical Implications of Spatial Complexity}\label{spatialcomplexity}

In this appendix, we give a brief account of the recent developments on the evolution of turbulent magnetic and velocity fields in electrically conducting fluids based on the concept of spatial complexity. Details can be found in \cite{JV2019}; \cite{JVV2019}; \cite{SecondJVV2019}.

The velocity field $\bf u$ at a fixed point $\bx$ in a river has a well-defined and definite direction if observed from a distant point; e.g., if the river flows from east to west, the velocity field will point from east to west. One may call this the large scale velocity field. However, as we approach the river and look at smaller and smaller scales around the point $\bx$, we see more complex motions in different directions---the direction of the small scale velocity field will not generally be from east to west. The large scale velocity at point $\bx$ can be defined as the coarse-grained field at a large scale $L$;

$${\bf{u}}_L ({\bf{x}}, t)=\int_V G_L({\bf{r}})  {\bf u}({\bf{x+r}}, t) d^3r, $$

which is the average velocity of a fluid parcel of size $L$ located at $\bx$. Here, $G_L({\bf r})=G({\bf r}/L)$ is a rapidly decaying function, e.g., $G_L({\bf r})\sim e^{-r^2/L^2}$, and thus the integral gets more contributions from points at a distance $\lesssim L$ from $\bx$ than distant points. The small scale field at $\bx$ is similarly defined as

$${\bf{u}}_l ({\bf{x}}, t)=\int_V G_l({\bf{r}})  {\bf u}({\bf{x+r}}, t) d^3r, $$

which is the velocity of a fluid parcel of size $l$ located at $\bx$. The above definitions of large and small scale velocities are examples of the mathematical objects called distributions.

How different are the directions of the small and large scale fields at point $\bf x$? Denote by $\theta$ the angle between these vectors, which is a function of space and time, i.e., $\theta=\theta({\bx}, t)$ and can be obtained easily using the inner product ${\bf u}_l.{\bf u}_L=u_l u_L\cos\theta$ or $\cos\theta=\hat{\bf u}_l.\hat{\bf u}_L$. Hence we can use this angle to quantify the difference between the directions of the large and small scale velocity fields at point $\bx$ and time $t$. The larger is the deviation angle $\theta$, the more complex the flow is at point $({\bx}, t)$. Moreover, if the flow is turbulent, it becomes a stochastic variable which measures the level of randomness in the velocity field at $({\bx}, t)$. Hence, a measure of spatial complexity, or level of randomness, of the velocity field $\bf u$ in a given volume can be defined as $S(t)={1\over 2}(\hat{\bf u}_l.\hat{\bf u}_L-1)_{rms}$. In what follows we formulate this idea in a more formal and general fashion which also introduces a measure for the energy of the field at different scales.

The quantity ${\bf u}_l.{\bf u}_L$ not only tells us how strong the small and large scale velocity fields are at $\bx$ (because it depends on the magnitudes $u_l$ and $u_L$) but also it tells us how parallel they are (because of its dependence on $\theta$). For a general vector field $\bf F$, we define 
\begin{eqnarray}\label{scalesplit}
\psi_{l,L}({\bf x}, t)&=&{1\over 2} \;{\bf{F}}_l({\bf{x}}, t){\bf{.F}}_L({\bf{x}}, t)\\\notag
&=&{1\over 2}\int_V d^nr\; G_l({\bf r})\int_{V} d^nr' \; G_L({\bf r}')\\\notag
&\times&{\bf F}({\bf x+r}, t).{\bf F}({\bf x+r'}, t),
\end{eqnarray}
as a generalized energy density at point $\bx$. The function $\psi_{l,L}: \mathbb{R}^n\times\mathbb{I}_t\rightarrow \mathbb{R}$ is in fact a scalar field\footnote{In general, for a complex field, one may consider $\psi_{l,L}({\bf{x}}, {\bf{R}}, t)={1\over 2}{\bf{F}}_l({\bf{x}}, t){\bf{.F}}^*_L({\bf{x+R}}, t)$, where ${\bf F}^*$ is the complex conjugate of $\bf F$. In this paper, we will only consider real fields and would take $\bf R=0$ hence $\psi_{l,L}({\bf{x}}, {\bf{R=0}}, t)=\psi_{l,L}({\bx}, t)$.} which noting that $\lim_{l\rightarrow 0} {\bf F}_l\rightarrow {\bf F}$, can be thought of as a generalization of the energy density;
 \footnote{For Lipschitz-continuous fields, we can consider an interesting limiting case as

\begin{equation}\notag
{\psi}_\infty({\bf{x}}, t):=\lim_{l\rightarrow 0} \lim_{L\rightarrow \infty} \psi_{l,L}({\bf{x}}, t)={1\over 2} \;\underbrace{  {\bf{F}}({\bf{x}}, t)}_\text{local\;field}.\underbrace{ \overline{\bf{F}}({\bf{x}}, t) }_\text{global\;field},
\end{equation}
where $\overline{\bf{F}}({\bx}, t)=\int_V  {\bf{F}}({\bx+\br}, t) (d^nr/V)$ is the (spatial) volume average of ${\bf{F}}$. The above expression follows from the properties of the kernel $G_l({\bf r})$.} $U({\bx}, t)$;

\begin{equation}
U({\bx}, t)={1\over 2}{\bf F}({\bx}, t).{\bf F}({\bx}, t)={1\over 2}F^2({\bx}, t).
\end{equation}

 It is also more convenient to consider direction and magnitude of the field separately by writing $\psi_{l, L}({\bf x}, t)=\phi_{l, L}({\bf x}, t)\chi_{l, L}({\bf x}, t)$ with two scalar fields

 \begin{equation}\label{phichi1}
\phi_{l,L} ({\bf{x}}, t)=\begin{cases}
\hat{\bf{F}}_l({\bf{x}}, t).\hat{{\bf{F}}}_L({\bf{x}}, t); \;\;\;\;\;\;F_l\neq 0\;\&\;F_L\neq0,\\
0;\;\;\;\;\;\;\;\;\;\;\;\;\;\;\;\;\;\;\;\;\;\;\;\;\;\;\;\;\;\;\;\;\;otherwise,
\end{cases}
\end{equation}
and

\begin{equation}\label{phichi2}
\chi _{l,L}({\bf{x}}, t)={1\over 2} F_l ({\bf{x}}, t) F_L({\bf{x}}, t).
\end{equation}

From equations (\ref{direction}) and (\ref{magnitude}), we realize that $\phi_{l, L}$ is related to the field's topology whereas $\chi_{l,L}$ is associated with the field's magnitude. In fact, for non-zero vectors $\hat{\bf{F}}_l$ and $\hat{{\bf{F}}}_L$, the scalar field $\phi_{l, L}$ is the cosine of the angle between two coarse-grained components of the vector field ${\bf{F}}({\bf{x}}, t)$ at different scales $l$ and $L$ at point $({\bf{x}}, t)$, i.e., $\phi_{l,L}=\cos\theta=\hat{\bf F}_l.\hat{\bf F}_L$. At any given point $({\bx}, t)$, this scalar field is simply what is known as the cosine similarity between two vectors\footnote{This is analogous to the Otsuka-Ochiai coefficient $\frac{|X \cap Y|}{\sqrt{|X| \times |Y|}}$ for two sets $X$ and $Y$, where $|.|$ denotes the number of elements, as the similarity measure. This measure in fact reduces to the cosine similarity if $X$ and $Y$ are bit vectors (i.e., maps from a set of integer numbers to the interval $[0, 1]$). }. This scalar field is called the topology field to imply its relationship with the vector field topology \cite{JV2019}. On the other hand, $\chi_{l,L}$ is in fact twice the geometric mean of the field energy densities at scales $l$ and $L$. In other words, we can write 
$$\chi_{l,L}= \sqrt{U_lU_L}$$
 where $U_l=F_l^2/2$ and similarly $U_L=F_l^2/2$. For simplicity, we will drop the index ${l, L}$ hereafter. 

Based on the quantities discussed above, we can now define the spatial complexity or self-entanglement (of order $p\in{\mathbb{N}}$) associated with the field ${\bf F}({\bf{x}}, t)$ as \footnote{The $L_p$ norm of ${\bf f}: {\cal{R}}^n\rightarrow {\cal{R}}^n$ is the mapping ${\bf f}\rightarrow ||{\bf{f}}||_p=[\int_V |{\bf f(x)} |^p (d^n x/V)]^{1/p}$. For $p=2$, $||f||_2=f_{rms}$ is the root-mean-square (rms) value of $\bf f$. For $p\leq q$, $||{\bf f}||_p\leq ||{\bf f}||_q$. Also $||{\bf f} ||_\infty=lim_{p\rightarrow\infty} ||{\bf f}||_p=||{\bf f}||_{\max}$. } 
\begin{equation}\label{stochasticity-rate0}
s_p (t) ={1\over 2} \Big|\Big| \phi - \parallel \phi \parallel_p\Big|\Big|_p.
\end{equation}
For instance, taking $p=2$, we find the second order self-entanglement;
\begin{equation}\notag
s_2 (t) ={1\over 2} \Big[ \phi -  \phi_{rms}\Big]_{rms}.
\end{equation}

This form resembles the definition of the conventional standard deviation corresponding to a random variable $X$; $\sigma_x=\langle (x-\langle x\rangle)^2 \rangle^{1/2}$ where $\langle.\rangle\equiv E[.]$ denotes the expected value calculated in the usual way using a given probability measure. In fact, in case we have a probability measure, one can use the standard deviation defined in the conventional way, that is

\begin{equation}\label{std}
s(t)={1\over 2}E\Big[\Big(\phi-E[\phi]\Big)^2\Big].
\end{equation}

Nevertheless, instead of probability measures, unlikely to be given in many real world applications, the definition (\ref{stochasticity-rate0}) is based only on spatial volume averages, which are easy to calculate in practice. In any case, although its numerical value will depend on the method used to calculate it, the concept of self-entanglement or spatial complexity will retain its meaning as a characteristic of the field. In fact, another even simpler measure can be defined in terms of the deviation of $\phi$ from unity rather than its ${\cal{L}}_p$-norm (or rms value $\phi_{rms})$ since for an unentangled, smooth field $|\phi|_p\simeq 1$. This form is more convenient since its time derivative is easy to work with, however, as statistical tools, they convey the same phenomenology, i.e., a generalization of deviation from the mean. Thus one can define the field's spatial complexity (self-entanglement) as \cite{JV2019}

\begin{eqnarray}\label{stochasticity-rate}
S_p (t) &=&{1\over 2} \parallel {\phi({\bf{x}}, t)} -1 \parallel_p,\\\notag
&=&{1\over 2} \Big[  \int_V \Big| \phi -1 \Big|^p {d^nx\over V}       \Big]^{1/p}.
\end{eqnarray}
The topological deformation (of order $p$) of the vector field ${\bf F}({\bf{x}}, t)$ is the rate of change of its spatial complexity with time  \cite{JV2019}
\begin{eqnarray}\label{topology-change}
T_p(t)&=& {\partial S_p(t)\over \partial t}\\\notag
&=&{S_p^{1-p}(t)\over 2^p } \int_V (\phi-1){\partial \phi\over\partial t} |\phi-1|^{p-2}\;{d^nx\over V}.
\end{eqnarray}

Using the formalism developed above, few other useful statistical measures can also be defined. For instance, the ($p$th order) cross energy corresponding to the vector field ${\bf F}$ is defined as

\begin{equation}
E_p(t)= \parallel{\chi}\parallel_p.
\end{equation}
Its time derivative corresponds to the dissipation rate;

\begin{equation}\label{Edissipation2}
D_p(t)= {\partial E_p(t)\over \partial t}=E_p^{1-p}(t)\int_V \chi {\partial\chi\over\partial t}\; |\chi|^{p-2}\;{d^nx\over V}.
\end{equation}

We can take $p=2$ for simplicity, therefore, the second order spatial complexity (self-entanglement) $S_2(t)$, topological deformation $T_2(t)$, cross energy $E_2(t)$, and dissipation $D_2(t)$ are respectively given by

\begin{equation}\label{formulae}
S_2(t)={1\over 2} (\phi-1  )_{rms},
\end{equation}
\begin{equation}
T_2(t)=  {1\over 4 S_2(t)} \int_V \;(\phi-1){\partial \phi\over \partial t}\; {d^nx\over V},
\end{equation}
\begin{equation}
E_2(t)=\chi_{rms},
\end{equation}
and
\begin{equation}
D_2(t)={1\over  E_2(t)} \int_V \chi \partial_t \chi{d^nx\over V}.
\end{equation}

One can use the renormalized evolution equation of the field, that is $\partial_t{\bf F}_l({\bf x}, t)={\bf f}_l( {\bf x}, t)$, to obtain explicit expressions for $T_2$ and $D_2$. The time derivative of the topology field $\phi({\bf{x}}, t)$ is given by
 
 \begin{eqnarray}\notag
{\partial \phi \over \partial t}&=&\Big[ {\partial_t {\bf{F}}_l\over F_l}.({\cal{I}}-\hat{\bf{F}}_l\hat{\bf{F}}_l)\Big] .\hat{\bf{F}}_L+
\Big[ {\partial_t {\bf{F}}_L\over F_l}.({\cal{I}}-\hat{\bf{F}}_L\hat{\bf{F}}_L)\Big] .\hat{\bf{F}}_l\\\label{localphi}
&=&\hat{\bf{F}}_L.\Big({\partial_t {\bf{F}}_l\over F_l}\Big)_{\perp{\bf{F}}_l}+
\hat{\bf{F}}_l.\Big({\partial_t {\bf{F}}_L\over F_L}\Big)_{\perp{\bf{F}}_L}.
\end{eqnarray}

Here, ${\cal{I}}={\cal{I}}_{n\times n}$ is the identity tensor and $(\;)_{\perp{\bf{F}}}$ represents the perpendicular component with respect to $\bf{F}$. We find

\begin{eqnarray}\notag
T_2(t)=&&{1\over 4S_2}  \int_V   \Big[\hat{\bf F}_l.\hat{\bf F}_L-1 \Big] \; \Big[  \hat{\bf{F}}_L. \Big({ {\bf f}_l \over F_l} \Big)_{\perp {\bf F}_l }  \\\label{T1}
&& + \hat{\bf{F}}_l. \Big({ {\bf f}_L\over F_l}  \Big)_{\perp{\bf F}_L }          \Big]  \;{d^nx\over V}.
\end{eqnarray}

The time evolution of the scalar field $\chi({\bf x}, t)$ can be similarly obtained,

\begin{equation}\label{Edissipation}
{\partial \chi\over \partial t}={1\over 2} F_l F_l \Big[ \Big( {\partial_t {\bf{F}}_L\over F_l } \Big)_{\parallel {\bf{F}}_L}+ \Big( {\partial_t {\bf{F}}_l\over F_l } \Big)_{\parallel {\bf{F}}_l}\Big].
\end{equation}
Here, $(.)_{\parallel{\bf{F}}}$ represents the parallel component with respect to $\bf{F}$. We have

\begin{eqnarray}\notag
D_2(t)&=&{1\over 4E_2}\int_V \Big[F_l^2\partial_t (F_l^2/2)+F_l^2\partial_t (F_l^2/2)    \Big]{d^nx\over V}\\\notag
 &=& {1\over E_2}\int_V \Big({F_l F_l\over 2}\Big)^2 \Big[ \Big( { {\bf{f}}_L\over F_l } \Big)_{\parallel {\bf{F}}_L}+ \Big( { {\bf{f}}_l\over F_l } \Big)_{\parallel {\bf{F}}_l}\Big]\;{d^nx\over V},\\\label{Edissipation30}
\end{eqnarray}
which is obviously related to the temporal changes in energy densities $F_l^2/2$ and $F_l^2/2$ at scales $l$ and $L$. This is in turn related to the parallel component of the renormalized evolution equation, i.e., $\partial_t {\bf F}_l={\bf f}_l$ (and $\partial_t {\bf F}_L={\bf f}_L$ at the scale $L$). For a physical application, see \cite{JV2019}; \cite{SecondJVV2019} and \cite{JVV2019}.

\section{Continuity of Field Lines}\label{fieldlinecontinuity}

The condition for the integral curve $\bxi_\bx$, which satisfies
eqs.(\ref{fieldline4}), to be continuous in time $t$ is given by $\lim_{\epsilon\rightarrow 0}|\bxi_\bx(s, t_0+\epsilon)-\bxi_\bx(s, t_0)|\rightarrow 0$ for arbitrary curve parameter $s$. We write 
\begin{eqnarray}\notag
&& \Big| \bxi_{\bx}(s, t_0+\epsilon)-\bxi_{\bx}(s, t_0)   \Big|\\\notag
&&= \Big|\int_0^s ds'\Big[  \hat{\bf F}(\bxi_\bx(s', t_0+\epsilon), t_0+\epsilon)-\hat{\bf F}(\bxi_\bx(s', t_0), t_0)  \Big] \Big|\\\notag
&&\leq \int_0^s ds' \Big|  \hat{\bf F}(\bxi_\bx(s', t_0+\epsilon), t_0+\epsilon)-\hat{\bf F}(\bxi_\bx(s', t_0), t_0)   \Big|.\\\notag
\end{eqnarray}

Assuming that $\hat{\bf F}$ is Lipschitz in spacetime position vector $\vec{x}=({\bx}, t)$, i.e., 

\begin{eqnarray}\notag
|\hat{\bf F}(\vec{x}_2)-\hat{\bf F}(\vec{x}_1)|&\leq& K_0 |\vec{x}_2-\vec{x}_2|\\\notag
&=&K_0\sqrt{ |{\bf x}_2-{\bf x}_1|^2 +|t_2-t_1|^2      },
\end{eqnarray}

 for some $K_0>0$, we can write

\begin{eqnarray}\notag
&& \Big| \bxi_{\bx}(s, t_0+\epsilon)-\bxi_{\bx}(s, t_0)   \Big|\\\notag
&&\leq K_0 \int_0^s ds'\sqrt{   \Big| \bxi_{\bx}(s', t_0+\epsilon)-\bxi_{\bx}(s', t_0)   \Big|^2+\epsilon^2      }.
\end{eqnarray}
Therefore, we find

\begin{eqnarray}\label{continuity1}
&& {\partial\over\partial s}\Big| \bxi_{\bx}(s, t_0+\epsilon)-\bxi_{\bx}(s, t_0)   \Big|\\\notag
&&\leq K_0 \sqrt{   \Big| \bxi_{\bx}(s, t_0+\epsilon)-\bxi_{\bx}(s, t_0)   \Big|^2+\epsilon^2      }\\\notag
&&\leq K_0\Big(\Big| \bxi_{\bx}(s, t_0+\epsilon)-\bxi_{\bx}(s, t_0)   \Big|+|\epsilon|\Big),
\end{eqnarray}
which implies 

\begin{equation}\notag
\Big| \bxi_{\bx}(s, t_0+\epsilon)-\bxi_{\bx}(s, t_0)   \Big|\leq |\epsilon| \Big( {e^{K_0 s}-1\over K_0} \Big)
\end{equation}

This is the inequality given by (\ref{continuity}).

  \section{Scalar Fields as Perfect Fluids}\label{Vectors}
In this appendix, we illustrate how a velocity field can be attributed to a given scalar field $\phi({\bf x}, t)$ represented as a perfect fluid.

 The equation of motion is obtained from an action, with potential $V(\phi)$:

\begin{eqnarray}\notag
S&=&-\int \sqrt{-g} d^4x {\cal L}(\phi, \partial_\mu\phi)\\\label{inflation110}
&=&-\int \sqrt{-g} d^4x \Big( {1\over 2} \partial^\mu \phi \partial_\mu \phi +V(\phi) \Big),
\end{eqnarray}
where $g$ is the determinant of the metric $g_{\mu\nu}$. The Lagrangian density is ${\cal L}=-\Big( {1\over 2} g_{\mu\nu}\partial^\mu \phi \partial^\nu \phi +V(\phi) \Big)$, and the Euler-Lagrange equations read $\partial^\mu{\partial {\cal L} \over \partial (\partial^\mu\phi)}={\partial{\cal L}\over \partial\phi} $, which give the equation of motion. For instance, with the quadratic potential $V(\phi)=m^2\phi^2/2$, we get the Klein-Gordon equation: $(-\partial_\mu\partial^\mu+m^2)\phi=0$. Note that here we use the metric sign convention $(-, +, +, +)$ and work in natural units thus $c\equiv 1\equiv \hbar$. 

The density and pressure corresponding to the scalar field $\phi$ is obtained by comparing its energy-momentum tensor to that of a perfect fluid:

$$\begin{cases}
T^{\mu\nu}=  {\cal L} g^{\mu\nu}+\partial^\mu\phi\partial^\nu\phi,\;\;\;\;\;\;\;\text{(scalar field)}\\
T^{\mu\nu}=P g^{\mu\nu}+(P+\rho)u^\mu u^\nu,\;\;\;\;\;\;\;\text{(perfect fluid)}\\
\end{cases}$$
with metric sign convention $-1=g_{\mu\nu}u^\mu u^\nu$. Here $u^\mu=dx^\mu/d\tau$ is the four-vector while $P$ and $\rho$ are the pressure and density of the fluid. We find the density, pressure and velocity field of the perfect fluid, respectively, as the following:
\begin{eqnarray}\notag
\rho=-{1\over 2} g^{\mu\nu}\partial_\mu\phi\partial_\nu\phi+V(\phi),
\end{eqnarray}

\begin{eqnarray}\notag
P=-{1\over 2} g^{\mu\nu}\partial_\mu\phi\partial_\nu\phi-V(\phi),\end{eqnarray}

and
\begin{eqnarray}\label{quintessence7}
u^\mu={\pm g^{\mu\nu} \partial_\nu \phi\over \sqrt{ -g^{\mu\nu} \partial_\mu \phi\partial_\nu \phi   } }.
\end{eqnarray}
 The sign of $u^\mu$ is arbitrary and independent of $T^{\mu\nu}$. We can choose it in a way that $u^0$ is positive for a positive $\dot\phi$, i.e., an increasing scalar field. We find
\begin{eqnarray}\notag
u^\mu&\equiv&\Big({dt\over d\tau}, {d{\bf x}\over d\tau}\Big)\\\notag
&=&{\pm 1 \over\sqrt{ -g^{\mu\nu} \partial_\mu \phi\partial_\nu \phi   }}\Big(\dot\phi , - g^{ij}{\partial \phi\over\partial x^i} \Big),
\end{eqnarray}

where $\dot\phi\equiv\partial \phi/\partial t$, and $g^{ij}$ is the purely spatial part of the metric $g^{\mu\nu}$. Noting that $u^\mu={dt\over d\tau}(1, {d{\bf x}\over dt})$, we find

$$\gamma\equiv {dt\over d\tau}={ \dot\phi \over\sqrt{ -g^{\mu\nu} \partial_\mu \phi\partial_\nu \phi   }},$$
and (if $\dot\phi\neq 0$)
\begin{eqnarray}\notag
{dx^j\over dt}=- g^{ij}{1\over\dot\phi} {\partial\phi\over\partial x^i}.\;\;\;\text{ (no summation)  }
\end{eqnarray}

Consequently, the ordinary $3$-velocity in a general curvilinear coordinates with (spatial) metric $g^{ij}=h_i^{-2} \delta_{ij}$ is

\begin{eqnarray}\notag
 v_j&=&h_j {dx^j\over dt}\\\notag
 &=&- {1\over \dot\phi}{\partial\phi\over h_j\partial x^j}.\;\;\;\text{(no summation)}
 \end{eqnarray}
Using the gradient operator $\nabla_i\equiv {\partial\over h_i\partial x^i}$ in the curvilinear spatial coordinates, we can finally write the $3$-velocity as

\begin{equation}\label{threevelocity}
{\bf v(x}, t)=- {\nabla\phi({\bf x}, t)\over \dot\phi({\bf x}, t)}.
\end{equation}

In conclusion, a real, inhomogeneous, time-dependent scalar field $\phi=\phi({\bf x}, t)$, represented as a perfect fluid, defines a vector field ${\bf v}({\bf x}, t)$ at any point in spacetime at which $\dot\phi\neq 0$. 

As an example, with the Robertson-Walker metric in a flat universe, $-d\tau^2=-dt^2+a^2(t) d{\bf x}^2$, with scale factor $a(t)$, the $3$-velocity defined by the inflaton field is given by   ${\bf v(x,} t)=- a^{-1}\nabla\phi/\dot\phi$. (Dividing the gradient operator by the factor $a(t)$ is due to the stretching of the physical coordinates as a result of the expansion of the universe: $\nabla_{physical}\equiv a^{-1}\nabla_{comoving}$. Equivalently, we can include this in the spatial part of the metric $g_{\mu\nu}$, i.e., in the definition of $h_i$'s in the Robertson-Walker metric.)

 It is easy to check that this velocity satisfies the general relativistic Euler equation. For instance, in a flat and expanding universe (i.e., using the Robertson-Walker metric with zero spatial curvature):
\begin{eqnarray}\notag
{\partial {\bf v}\over \partial t}+a^{-1}{\bf v}.\nabla{\bf v}&=&-{1-v^2\over P+\rho}\Big( a^{-1} \nabla P+{\bf v}{\partial P\over \partial t} \Big)\\\notag
&&-{\dot a\over a}{\bf v}(1-v^2),
\end{eqnarray}
with $v\equiv |\bf v|$. For a non-expanding fluid, $a(t)\equiv 1$, and the Hubble parameter vanishes $H\equiv \dot a/a=0$. Thus, the $3$-velocity given by eq.(\ref{threevelocity}) satisfies the familiar special relativistic Euler equation:
\begin{eqnarray}\notag
{\partial {\bf v}\over \partial t}+{\bf v}.\nabla{\bf v}&=&-{1-v^2\over P+\rho}\Big( \nabla P+{\bf v}{\partial P\over \partial t} \Big),
\end{eqnarray}
which is eq.(\ref{Euler10}).

In the end, we note that the velocity given by eq.(\ref{threevelocity}) is in fact quite general and can also be obtained in a rather informal way. The simplest vector associated with a scalar field is its gradient, e.g., in electrostatics, the electric field is proportional to the gradient of the scalar potential; ${\bf E}=-\nabla\Phi$. We may in fact represent the scalar field as a fluid whose velocity field is proportional to the field gradient. In a relativistic setup, one can define the fluid's four-velocity as $U^\mu\equiv (\partial_t\Phi, -\nabla\Phi)$, but the relativistic relation $U^\mu U_\mu=-1$ would require then $-(\partial_t\Phi)^2+|\nabla\Phi|^2=-1$, which is an additional and undesired constraint. A simple solution is to normalize the velocity field and re-define it as 
 
\begin{eqnarray}\notag
U^\mu&=&{1\over \sqrt{(\partial_t\Phi)^2-|\nabla\Phi|^2}}(\partial_t\Phi, -\nabla\Phi)\\\notag
&=&{ {\dot\Phi\over |\dot\Phi|} \over \sqrt{1-|{\nabla\Phi\over\dot\Phi}    |^2}        }(1, -{\nabla\Phi\over\dot\Phi}).\end{eqnarray}

 which satisfies $U^\mu U_\mu=-1$ without setting any constraint on the field $\Phi$ provided that $\dot\Phi({\bf x}, t)\equiv \partial_t\Phi({\bf x}, t)\neq 0$. This leads to eq.(\ref{threevelocity}). The Lorentz factor is defined as $\gamma\equiv{dt\over d\tau}= \pm (\dot\Phi/|\dot\Phi |)(1-|{\bf v}|^2)^{-1/2}$ with plus (minus) sign chosen for $\dot\Phi>0$ ($\dot\Phi<0$). Therefore, $\gamma= (1-|{\bf v}|^2)^{-1/2}$ as desired.

\bibliographystyle{apsrev4-2}
\bibliography{VectorFields7}

\end{document}